# The Collaborative Critique: An Inspection Method for Expert Evaluation of User Interfaces

Wendy Lucas, Tamara Babaian
CIS Department, Bentley University, Waltham, MA, 02145, USA
*{wlucas,tbabaian}@bentley.edu*

March 5, 2015

**Abstract**

The development of the collaborative critique (CC) usability inspection method was driven by the need for systems, ranging from the simple to the complex, to lessen the workload on their users by being better collaborative partners. In the collaborative view of system-user interaction, the role of the system is expanded to that of a partner that must help its users achieve their goals. This perspective shifts the dynamic from focusing on the user's ability to make effective and efficient use of the system to the system's usefulness within a given task, user, and interaction context, thereby reducing the user's cognitive and physical efforts. The CC method consists of a walkthrough evaluation in which the questions answered at each step of the interaction assess the collaborative strength of the system. This method is further distinguished by the inclusion of questions that access the behavior of the system in error situations, which are known to be particularly burdensome to users of complex systems. In this paper, we provide a detailed description of the CC method and its application and demonstrate its usefulness in evaluating an interface. The results of a laboratory evaluation show that the method performed well in identifying the problems experienced by users across all severity levels and predicted almost all of the problems associated with error situations.

## 1 Introduction

The entrenchment of websites and other software applications in daily life has been accompanied by significant strides in both designing for and evaluating usability. Conquering the usability of complex systems, however, remains the "next frontier" [1, p. xiii]. System complexity arises from the underlying complexity of the tasks being supported [2], which may have indeterminate goals and criteria for completion, require higher-order cognitive skills and the integration of knowledge from different areas, and necessitate advanced learning and instruction [3]. The need to support a broad range of users with varying levels of experience performing a variety of tasks contributes to the difficulty of designing complex systems that are both usable and useful [4, 5].

Achieving better performance in complex systems is not simply a matter of deciding which tasks can be delegated to the computer and to what extent, and then developing algorithms for automating those tasks [6]. As research has shown [7, 8], the design of system-user interactions is a delicate balancing act in which the appropriate separation of responsibilities between a human and a computer is just the starting point. Researchers (e.g. [9, 10, 8]) have argued that the ultimate goal of system design should be to create a means for true and effective *collaboration* between a user and a computer system, where collaboration means a group of partners (in our case, a person and a computer system) working together with a commitment to achieving a shared goal. Such collaboration can take on different forms, but it always requires addressing many concerns beyond who does what in working towards the goal. For successful collaboration, there must be clear and effective communication between the parties about the parameters of the task, timely sharing of relevant information, and performing of helpful actions that support each other's efforts. These requirements motivated the development of a usability evaluation method that uses human-computer collaboration as the underlying paradigm, focusing on the strength of the system interface in collaborating, rather than merely interacting, with the user.

The collaborative critique (CC) usability inspection method was developed for guiding the design of systems that act as partners to their users, thus reducing their cognitive and physical efforts. It focuses on evaluating the extent to which a system collaborates with its users in achieving their goals. Babaian et al. [11] presented an initial laboratory evaluation validating the usefulness of the CC method. The findings showed that the method uniformly detected problems across the severity spectrum, revealed systematic as well as localized usability problems, performed especially well in error situations, and resulted in the reporting of very few false positives. These results indicate that the CC method mitigates many of the shortcomings reported in studies of other usability inspection techniques, such as heuristic evaluation (HE), heuristic walkthrough (HW), and cognitive walkthrough (CW) (see [12] for a review of these methods).

The CC inspection is conducted as a walkthrough in which evaluators perform realistic tasks with a system, as specified by an application scenario, and respond to a set of questions at each step. The human-computer collaboration paradigm [13], which is grounded in collaboration theory [14, 15], provided the framework for formulating the CC questions.

Recent studies have revealed that usability inspection methods in general fail to measure the usefulness of the overall system, becoming increasingly inadequate as system complexity increases [16, 17]. Applying the collaborative view to evaluating system-user interactions shifts the focus from analyzing individual features and functions to assessing the integrated use of multiple features required for achieving a comprehensive goal. Being a collaborative partner requires that the system does its part by sharing information and adjusting its behavior based on its knowledge and awareness of the user, the context of the interaction, and its own functionality. This perspective is realized in CC ques-tions that focus attention on the system's usefulness, including its effectiveness at providing guidance in performing tasks, reducing the amount of user effort required, and communicating information concerning the progress that has been made and what remains to be done.

Another important consideration for complex system usability is how helpful the system is in terms of both preventing errors and assisting with recovery when an error does occur [18]. Studies have shown that a lack of support in error situations can lead to significant difficulties in using complex systems, leaving users feeling lost and confused [19, 20]. It is therefore imperative for usability inspection methods to explicitly consider error situations if those methods are to be used effectively for evaluating complex systems. The CC method includes questions that focus on the helpfulness of the system in error situations, which are only answered when an error condition arises. The general questions concerning the utility of the system in guiding, communicating, and reducing user effort are also useful in revealing the presence or absence of supportive features and interventions that help prevent errors.

Lastly, studies have shown that the quality of usability evaluations is affected by the quality of the resources provided to the evaluators [21, 22]. In a CC evaluation, particular attention is paid to selecting a realistic task for evaluation, following a well-defined procedure, and providing evaluators with resources on the user and the task.

In this paper, we first discuss popular walkthrough methods and their limitations, which contributed to the need for developing another expert inspection technique. Next, we describe in greater depth the theoretical foundations for the CC method and present the evaluation questions that were formulated from this theory. We then detail the evaluation procedures and provide an example of a CC walkthrough inspection that demonstrates its application in practice. A summary of an evaluation of this method is then presented. This is followed by a discussion of the strengths and limitations of the CC method versus other inspection techniques. We conclude with directions for future research.

## 2 Review of Related Methods

Usability inspection methods have gained popularity as cost-effective alternatives to user testing due to their reliance on evaluators instead of users [23, 12]. In this section, we describe four inspection techniques and discuss their strengths and limitations. Three of these, namely cognitive walkthrough (CW), heuristic evaluation (HE), and heuristic walkthrough (HW), are well-established evaluation methods [12]. The fourth, metaphors of thinking (MOT), is a more recent psychology-based method that aims to address some of the shortcomings of other inspection techniques [24]. As noted by Cockton et al. [24], it is very difficult to rank usability inspection methods or make firm recommendations on the superiority of one versus another because their performance is highly dependent on how a particular method is configured and used in practice; practitioners must "work to get methods to work" [24, p. 1280]. In light of this difficulty, we present a discussion from the literature contrasting these four methods.

Grounded in the model of learning by exploration (CE+) [25], cognitive walkthrough was originally designed for evaluating simple *Walk Up and Use* interfaces for their learnability [26, 27, 28]. Subsequent revisions have both simplified its application and extended it to include more complex tasks by providing evaluators with step-by-step instructions on the action sequence to be performed [24]. During an inspection, evaluators attempt to determine if the user would follow the correct action. In making this assessment, the four CW questions, which focus on four aspects of learnability (see Figure 5 in Section 7), are considered [29]. The first of these questions asks if the user will try to achieve the right effect, while the other three questions assess if the user will choose the appropriate action and, if so, see that progress is being made on the task. For each correct action, the evaluators must tell a credible story of either success or failure, with the latter indicating a potential usability problem.

Most of the studies of CW have involved earlier versions and are therefore not discussed here. In evaluating the later, four-question approach, Lewis and Wharton [28] found that CW identified unique problems not identified by other methods. In reporting on two case studies, Jacobsen and John [30] found that CW did not predict usability problems related to missing functionality or error recovery.

In a heuristic evaluation [31, 32], evaluators inspect part of an interface in light of a set of heuristics that were derived empirically by Nielsen (see his revised set of heuristics in Figure 1 [33]). While evaluators are provided with these "rules of thumb," they are given little guidance concerning the users themselves or the tasks they perform. In [32], it is recommended that HE be performed as at least a two-pass process, with evaluators getting the feel for the interface in the first pass and then focusing on elements of the system in the second pass. This approach is very similar to usability walkthroughs [34], a two-pass evaluation method utilizing a set of evaluation guidelines in which the first pass involves self-guided exploration of the interface and the second involves the use of prescribed scenarios.

The heuristic walkthrough method [35] combines features from HE, CW, and usability walkthroughs. An HW is also a two-pass process but is more structured than the HE approach, as the evaluators have a prioritized list of user tasks in addition to a list of heuristics and "thought-focusing" questions [35]. HW includes both free-form and task-based evaluations. In the first pass, evaluators are guided by the questions as they explore tasks from a prioritized list in whatever order they choose. The second pass is guided by the list of usability heuristics, with evaluators able to explore the system. Problems are documented and assigned severity ratings by the evaluators during both passes, with ratings being consolidated into one severity rating per problem post evaluation.

In a study by Sears [35] comparing HW to CW and HE, heuristic walkthroughs were shown to identify significantly more intermediate and minor problems than CW. HW found more severe usability

1. Visibility of system status
2. Match between system and the real world
3. User control and freedom
4. Consistency and standards
5. Error prevention
6. Recognition rather than recall
7. Flexibility and efficiency of use
8. Aesthetic and minimalist design
9. Help users recognize, diagnose, and recover from errors
10. Help and documentation

Figure 1: Nielsen's ten heuristics.

problems than CW when one or two evaluators were involved, but that difference went away when more evaluators participated in the assessment. Both heuristic and cognitive walkthroughs yielded fewer false positives than heuristic evaluations. This can be explained by the lack of task context in HE, which means that the relevance of heuristic violations to the task is not assessed.

The fourth inspection method considered here identifies usability problems as violations of metaphors of thinking (MOT) [36, 37]. The aim of the five metaphors used by this method is to stimulate inspection and aid both evaluators and designers by focusing attention on the user's mental activity and how it is taken into account in the user interface. These metaphors concern habit, the stream of thought, awareness and associations, the relation between utterances and thought, and knowing. In a MOT inspection, the evaluators define representative tasks and walk through the interface with those tasks in mind. Usability problems are typically identified and noted through consideration of the metaphors and associated questions [37]. In comparison to HE, the MOT guidelines are much more complex for the evaluators to interpret than straightforward heuristics that focus on specific usability factors.

According to [37], the weaknesses of other inspection methods that MOT aims to address through the use of metaphors are (1) the lack of explicit consideration of the users' thinking (the authors of [37] note that CW has less emphasis on the psychological basis of thinking), and (2) limitations on the usefulness of those methods to particular devices and use contexts. Their experimental evaluations demonstrated that MOT found more problems than either CW or HE, and that those problems were more complex and likely to persist for expert users. While CW, HE, and HW have been widely used in practice, MOT is a newer technique so less is known about its relative strengths and limitations.

In Section 7 of this paper, we discuss how the CC method differs from those examined above and summarize its strengths and weaknesses. First, however, we must describe this method, its background and application, and results from an initial evaluation.

# 3 Theoretical Background and CC Questions

Norman's model of the human action cycle [38] identifies three continuously repeated stages of human-computer interaction:

1. the goal formation stage, during which a person is using his/her mental model of the task and the context of interaction to formulate the next goal,
2. the execution stage, during which the person translates his/her goal into concrete steps involving the use of the system and executes those steps, and
3. the evaluation stage, during which the person assesses the outcome of executing the steps and further evaluates the state of the world with respect to the goal.

This model has been successfully used, either explicitly or implicitly, in most software interface design and evaluation methods to date. With our growing dependence on computer systems that are increasingly sophisticated, it is advantageous to bring the designed behaviors of the systems into the picture [39]. Norman writes in "The Design of Future Things" ([39], page 4):

> *"As our technology becomes more powerful, its failure in terms of collaboration and communication becomes more critical."*

Collaboration theory, with its focus on what is required for a successful and productive partnership, serves as a natural guide for including the system's behavior in a systematic way in the modeling, design and evaluation of human-computer interaction. Collaboration theory can help focus the designers' attention on the aspects of this interaction that make for an effective human-computer partnership, where the system acts in a way that is consistent with our notion of a committed co-worker or assistant [9]. Collaboration theory defines the principles that distinguish collaborative activity from merely working together. It has been applied by researchers to inform novel, smart system-user interaction designs [40, 41, 42]. Among the widely available systems, the Java development component of the `Eclipse` [43] integrated development environment is an exemplar of a collaborative design.

As summarized by Terveen [13], all theoretical accounts of collaboration (often referred to as *collaborative activity*, or *joint activity*) emphasize the following key attributes:

1. A *goal* of collaboration that is shared by all participants. This goal may not be completely specified at the start and may be refined in the process of collaboration.
2. *Plans* for achieving the goal. The collaborating partners must agree on the division of tasks between the parties and plans for coordinating their individual and group activities. As with the goal, plans are typically incompletely specified in the beginning and evolve with time.
3. Partners must maintain a *shared context* of their interaction including information on their progress towards their goal.

4. *Communication* is essential for information sharing while developing the goal and plans and for maintaining the shared context.

5. *Learning and Adaptation* as well as other kinds of *helpful behaviors* that arise from the parties' commitment to maximizing the chances for the success of the joint activity.

In addition, Bratman [14] notes that critical to collaboration is the *willingness of the participants to help each other* in case one of them is having a problem fulfilling their assigned part of the joint activity. This requirement is motivated by the individual's commitment to the successful completion of the task.

Let us now consider what these collaboration-related concepts mean in the context of system-user interaction and how they are typically handled within the traditional design paradigm. Each system is designed to work with the user on a range of goals. This range, along with the plans that a user and computer may enact in pursuit of a goal, depend on the functional capabilities of the system. Thus, the goals and the associated plans are bounded by the system's functionality. Designing for effective collaboration must, therefore, start with the identification of the range of goals and the ways in which a computer system may contribute to an effective sharing of the workload. This is where traditional design and evaluation methods typically focus most of their attention [44].

Along with the system's functionality, traditional methods concentrate on optimizing the system's front end as a means for the users to communicate their immediate goals and receive system feedback on the requested action. Users specify their goals by picking the appropriate options among those provided by the system's interface, and the system typically responds by informing the user about the results of taking the requested action. Focusing design and evaluation around each individual iteration of Norman's action cycle in such a way works reasonably well for simple tasks. However, to achieve greater effectiveness when using a system on a comprehensive task consisting of heterogeneous parts and multiple possible courses of action, the design must recognize that each individual iteration of the action cycle is usually part of a broader task process [4].

Consider the system-user interaction over the scope of a complex task. As the user is working with the system, the original goal gets decomposed into lower-level subgoals and plans for accomplishing them. These original subgoals get replaced with new ones as some of the subgoals are achieved or reformulated. In current practice, it is often assumed that the user will know and remember all essential task and system interface details required to accomplish even the most complex goals. The user is expected to know the sequence of required steps, the parameters controlling the execution of the task, and how to navigate to the appropriate part of the system interface and specify those parameters. The amount of resources spent on training and continuously supporting users in the workspace demonstrates time and again, however, that these expectations on the user's knowledge and abilities are unrealistic except for when dealing with interfaces designed for short and trivial tasks.

The collaboration view of human-computer interaction requires the system be designed to share the responsibility for efficient task completion by considering the context of the task, communicating relevant information, using its available resources to provide helpful assistance, and guiding the user, as necessary. In a true collaboration, the ongoing goal and plan refinements require input and information sharing from all participants for maximum success [15]. In order for this information exchange to be efficient, the participants must take into consideration the context of their interactions [45, 4]. A system that maintains the shared context helps ensure that system-to-user communication is relevant and thus well understood by the user, thereby minimizing the need for the human to provide details that are directly implied by the context. Context awareness (which is closely related to *situation awareness* [46]) also enables the system to recognize and perform helpful actions, such as sorting the results of a search for a movie theater based on the proximity to the user. Systems that utilize contextual information, learn from past experiences, and adapt to best fit the task and the users' needs are perceived as smart and useful. On the opposite end of the spectrum are systems that require the same or obvious information be entered time after time. This latter kind of design is clearly non-collaborative, since such systems are not built to use the information at their disposal in a maximally helpful way, violating the required commitments to the activity and to supporting their users.

In designing for effective system-user collaboration, one does not need to emulate human behavior in the system. Given the mostly orthogonal strengths of people versus computers, the system's design can follow the *human-complementary* [13, 9] approach. With this approach, tasks are allocated between the human and the computer according to the natural strengths of each; communication between the system and the user is also designed to leverage those strengths without having to rely on human discourse.

While examining the communication requirements at the basic level, user interface inspection methods typically do not address (1) how helpful the interface is in supporting the user in assessing the state of the task and determining the optimal way to proceed, (2) how efficiently it makes use of the task and user context of the interaction in presenting the user with options and minimizing the physical effort required, and (3) whether the system acts as a helpful assistant when an error situation occurs or a user is otherwise unable to proceed effectively. When people face a simple and familiar task, these concerns are mitigated by their background knowledge and capacity to learn and adapt to the system. However, for systems that embody a broad and diverse scope of complex or unfamiliar goals, it becomes imperative on the interfaces to inform and guide their users through the space of possible plans. These considerations motivated the development of the collaborative critique.

### 3.1 Collaborative Critique Evaluation Questions

Collaborative critique is a walkthrough method in which the evaluation is performed via the process of answering a predefined set of questions while performing a specified task. The CC questions focus on the system's usefulness throughout the different stages of Norman's cycle of user interaction. In this section we introduce the questions and discuss their theoretic underpinnings.

**Question 1: Will the user find the options for what she wants to do in the current screen?**

This question is assessing if the user, who may or may not know exactly what she wants to accomplish, will be able to find in the currently displayed interface an option that matches her mental model for what should be done next. A positive response indicates that the option is available and easily identifiable within the screen presented by the system. A negative response could result from a number of situations, such as:

- The user is too confused to understand what she wants to do next.
- The user knows what she wants to do, but there is no such option readily available in the current screen.
- The user knows what she wants to do, and the option is available, but the user cannot associate the option's label with that action.

Thus, Question 1 requires that evaluators take a look at the current screen and assess (1) how easy it is to relate one's understanding of the current state of the task in progress to what is depicted on the screen, and (2) whether the content of the screen supports the process of moving towards the next goal or the next step in the plan.

Questions 2a and 2b further refine the assessment of the match between the user's understanding of the possible next steps within the given task context and the related communication from the system.

**Question 2a: For the user to figure out what to do now, how much exploration is involved? Considering users with a range of experiences, answer with a number 1..5, where**

**1: most users will know what to do right away**
**2: some users will have to explore to figure out what to do**
**3: most users will have to explore to figure out what to do**
**4: most users will have to explore and some will be unable to figure out what to do**
**5: most users will have to explore and most will be unable to figure out what to do**

**Question 2b: For the user to figure out what to do now, how much confusion is involved? Considering users with a range of experiences, answer with a number 1..5, where**

**1: most users won't be confused**

**2: some users will be somewhat confused**

**3: most users will be somewhat confused**

**4: some users will be very confused**

**5: most users will be very confused**

Question 2a refers to the need for exploration of the current interface. By exploration, we mean the user's visual inspection of the current screen and the physical interaction with the interface controls in order to find the right option or value, or to form a plan to achieve the next goal.

Question 2b refers to the user's confusion at the present point. This question is assessing the user's understanding of what is going on in the current interaction with the system, what needs to be done now, and how to do it.

The evaluators may provide a comment along with the numeric value regarding a category of users that might experience difficulty. Even without such a comment, Questions 3a and 3b below are intended to further highlight the reasons for the reported levels of user confusion and exploration and to point out what can be improved in the design of the interaction.

**Question 3: Is the system using knowledge of the task in general, the current user, and the context of the current action to the fullest extent in order to:**

**a. appropriately guide the user?**

**b. reduce the effort involved in user input?**

These questions assess system behavior from the perspective of the requirements on each party to share knowledge and to act for the optimal outcome of the joint action. This kind of knowledge sharing on the part of the system is the subject of Question 3a. The guidance mentioned in this question encompasses a variety of system behaviors aimed at informing the user about the important parameters and possible courses of action in accomplishing the task. Examples of task-related guidance include:

- distinguishing between required parameters and optional ones,
- displaying instructions when the mouse is hovering over an interface item,
- offering a tutorial when the user has just entered the interface for the first time, or
- providing a view of tasks related to the current step in the process.

As mentioned earlier, such guidance is especially important in complex systems in which the user cannot be expected to know the multitude of available options and their impact on the task at hand.

A positive response to Question 3a indicates that the system is offering sufficient guidance for the user to be effective and efficient in deciding what steps to undertake and how to do so. When the next step is obvious without any additional information, the answer to this question should be NA (Not Applicable). A negative response to this question indicates that there is a need for helping the user in deciding how to proceed in the current situation, yet the system is not providing adequate information.

In a productive collaboration, the participants act in a thoughtful way, selecting appropriate choices for planned actions and action parameters in order to minimize unnecessary communication and physical effort. Furthermore, collaborators shape their individual behavior based on what they know about each other's strengths, what they've learned about each other by working together in the past, and by keeping aware of the relevant aspects of each other's ongoing efforts for the sake of the efficiency of the joint action [14]. Question 3b addresses these concerns by asking if the system is leveraging its knowledge of the task, context, and available resources *fully* to minimize the efforts of the human partner. This kind of helpful assistance in reducing the user's mental and physical efforts is the subject of this question. It distinguishes a "lazy" system, which asks irrelevant or obvious questions, "hides" pertinent information for the user to uncover, or leaves it to the user to filter out the useful choices, from a responsible collaborator whose reasoning, proactive behavior, and attention to relevant details result in useful suggestions that save users a lot of time and lead to greater user satisfaction. An NA response to this question is appropriate when there is no reasonable way for the system to reduce the user's efforts.

It must be emphasized that in Questions 3a and 3b, the evaluators have to assess if the system is using the contextual information and its knowledge of the task and the user to the maximum extent in presenting appropriate guidance and acting helpfully. For example, imagine a user who is reading an email message that includes the date and time for a meeting. When this user clicks on the calendar (maintained in the same system) in order to create an entry for the meeting referenced in the message, a helpful system will detect the date mentioned in the email and will present the user with the option of opening the calendar directly to that date and time and linking the calendar event to that email. Another example illustrating the importance of context-aware action as opposed to a generic display of all options regardless of their relevance: before suggesting a change of file permissions from "read only" to "read and write," the system must be sure that the latter option is available given the user's authorization level with respect to the file. A system offering useless suggestions and comments is not providing a form of guidance; it is creating a distraction that may take the user in a wrong direction and cost time.

**Question 4. After execution of the current action, will the user understand**

**a. what progress has been made so far toward completing the overall task?**

**b. what remains to be done in order to complete the overall task?**

Question 4 assesses if the system provides sufficient support during Norman's *evaluation* stage. From the collaboration standpoint, it is essential for the system to inform the user about what has been successfully accomplished in order for the user to move on to the next step. Question 4a asks if, given the interactions with the system in working on the task, the user understands what has been done and is able to keep track of where she is in the process. In determining their progress, users typically rely on their general knowledge of the task plus any cues displayed by the system to mark the accomplishment of important parts of that task.

A positive response to Question 4a indicates that, given the overall task, the user is sufficiently confident in her understanding of what has already been accomplished toward completing it. A negative response indicates that the user is unsure of what has been accomplished.

For example, consider the simple task of sending an email with an attachment. If the user sees the data she entered into the required fields, the text of the message she entered, and an icon with the name of the attached file, that should be sufficient for the user to feel confident that she understands what has been done and that the only remaining action is to send the message.

For more protracted, complex tasks, users may need to rely on some kind of a visual representation of the task components or other indications of where they are in the process. For example, consider the task of editing a digital video, during which an appropriate set of image transformations must be selected and applied until some kind of optimal outcome is achieved. To help the user determine how to proceed, it would be helpful for the system to display the transformations that have been applied already, perhaps including some thumbnail representation of the result of each stage for a quick review.

Question 4b is similar to 4a but assesses the user's understanding of what still remains to be done. A positive response indicates that the user knows exactly what is left to do. A negative response indicates that the user is unsure of what remains to be done. For example, in a task involving a sequence of data entry forms, the user may not be aware of how many more fields or screens the system will present before the task is completed. For an open-ended task without a clear goal, the system could inform the user of the commonly performed next steps or related tasks. When this kind of information is not clearly identifiable, an NA response is appropriate.

Questions 5, 6, and 7 must be answered in the case of an error situation. They are motivated by Bratman's requirement on collaborators to step in when a breakdown in a partner's actions occurs and they can be of help. This is one of the key properties distinguishing individuals merely working together from collaborating with each other and thus acting on their commitment to the success of the joint activity. As in Question 3, Questions 5 to 7 require the system be an active participant in diagnosing the problem and working to achieve a quick and effective resolution, based on its abilities and knowledge. This stance is contrasted by a more prevalent design in which the system acts as a passive bystander, leaving the job of diagnosing and fixing the problem to the user, despite the system actually having full access to the diagnostic information and possible corrective actions.

**Question 5: Does the system display information that clearly explains the problem to the user?**

A positive response to this question indicates that, in responding to an error situation, the system communicates to the user what has gone wrong in a clear and understandable way. A negative response indicates the error message is cryptic or gives insufficient details for the user to understand what went wrong.

For example, consider an error reported by the system when the user is trying to create a new file with the name of an existing one. A positive response is appropriate if the system reports that there already is a file with the same name and, therefore, it cannot create the requested file. A negative response is appropriate in the situation where the system reports something like "Error creating file" or "Unable to create file" without disclosing the reason.

**Question 6: Does the system present steps the user can take for possible corrective actions?**

Consider the file example from the previous question. A positive response to Question 6 is appropriate if, along with informing the user that a file with that name already exists, the system suggests that to eliminate the error the user might: (a) delete the existing file, (b) rename the existing file, (c) choose an alternative name for the new file, or (d) choose an alternative location for the new file. A negative response is appropriate when no useful suggestions are given to the user or the ones given are too vague to be helpful.

**Question 7. Does the system present an easy way to take corrective actions?**

Continuing with the same example, a positive response to Question 7 indicates that, along with informing the user that a file with that name already exists, the system includes executable options for actions (a) through (d) above. Upon the user selecting one of those options, the system would execute it, presenting a convenient interface if any user input is necessary. A negative response is appropriate when no executable options for eliminating the problem are provided.

# 4 CC Walkthrough Procedure

In this section, we describe the procedure for conducting a CC usability inspection. The CC walkthrough is performed by one or more evaluators who follow a specified task sequence and, for each action in that sequence, answer the questions reviewed in Section 3.1. As with the Cognitive Walkthrough method [47], there are two phases to the CC procedure: the preparation of the resources

**User-related:**

- A description of a typical user who performs the task.

**Task-related:**

- Training material on the task itself and how it is implemented within the system.
- A description of the task to be performed in the CC evaluation.
- The sequence of correct actions to be performed for the task.
- Identification of a domain expert(s) who will be available during the evaluation.

**Collaborative Critique method-related:**

- An electronic spreadsheet for recording answers to the CC questions for each action in the sequence.

Figure 2: Sample of resources for CC preparation.

required for performing an evaluation, and the evaluation itself. Each of these components are presented below.

## 4.1 Preparation

The first step in the preparation is to ***select one or more tasks for evaluation***. While the selection of a task will depend on the goal of the evaluation and is within the purview of the investigator, it must be a *realistic task* that is actually performed by users in practice for the findings to be relevant and useful. Complex tasks characteristically consist of multiple subtasks, each of which are comprised of steps, a. k. a. actions, that the user must perform. The user may also need to access interfaces and perform actions that are outside of the strict definition of what constitutes a particular task. For example, system administrators responsible for assigning authorizations to users may also need to perform tasks associated with user creation and role definition. A task selected for evaluation must therefore *incorporate features and functions that require the user to access multiple interfaces within the system.*

An additional consideration in selecting a task is the importance of evaluating system support in error situations. A criterion in task section should therefore be the *likelihood of encountering error conditions* during evaluation, either because the task naturally leads to such situations or because it allows for the "seeding" of actions resulting in errors.

Once the set of tasks has been identified, supporting resources will be needed for providing evaluators with the requisite knowledge and information for conducting the CC inspection. This includes materials related to the user, the task(s) to be performed, and the CC method itself. Figure 2 provides a sample list of the resources required for an evaluation. While more than one task can be selected, we refer to the evaluation of a single task in the figure and in the following discussion for the sake of clarity.

**Task Description: Purchase Order**

**Create a Standard Purchase Order**

*Create a purchase order– Vendor/Supplying Plant Known (transaction code ME21N)*

Vendor: Dell Computers

Purchasing Organization number 3000 (IDES USA)
Purchasing Group: P99 IDES SG G99 (Ext)
Company Code 3000 (IDES US INC)
Materials:
- Dell Latitude D600-2
- Telephone AT Company Model AT-50

Quantity:
- 3 for D600-2
- 10 for AT-50

Figure 3: Snippet of a sample task description.

#### 4.1.1 User-Related

Throughout a CC inspection, evaluators must put themselves in the place of a user. This necessitates the preparation of a ***description of the typical user***. This profile should include any relevant information about the user's role, training, administrative rights, and responsibilities as they relate to performing the selected task. It should also specify assumptions about the user's knowledge, including familiarity with the task and experience level and what the user should and should not be expected to know.

#### 4.1.2 Task-Related

To perform an evaluation of a complex, as opposed to a walk-up-and-use, application, the evaluators will likely need to be provided with ***training material on the task and its implementation within the system***. That material can take many forms, including written documentation, use case diagrams, etc. An overview of the task and its purpose, along with task-specific terminology that would be known to the profiled user, should be defined. A procedural description for each subtask, including the required steps or actions, will also need to be provided for complex tasks requiring prior user experience or training. Disclosure of the particular parameters of the evaluation to be conducted, however, such as the actual data that will be used in the evaluation task, should be avoided in order to reduce the likelihood of introducing evaluator bias.

A ***task description*** is required for each task to be performed in conducting the CC inspection. This document describes the overall task and the context in which it is performed, the goal of each subtask, the intended effects of the actions to be performed, and the required input data. Figure 3 shows a partial task description for a sample task involving the creation of a purchase order in an Enterprise Resource Planning (ERP) system, which is used for managing organizational processes and resources on an enterprise-wide basis and is an acknowledged example of a complex software application [2, 20].

While the task description provides evaluators with a general idea of what they will be doing, the ***sequence of correct actions*** is required for specifying in detail the steps to be enacted by the evaluators in performing the task. Steps are organized by subtasks into actions and, when appropriate, subactions. For example, the action of entering a data value into a text box may include the subactions of accessing a search interface, entering a search term, and selecting a data item from the results returned. The sequence of correct actions should generally adhere to the system-prescribed means for performing the task, with the term "correct" referring to the actions that the evaluators should be performing rather than the outcome of those actions. For example, the "correct sequence" may specify selection of the first result returned from a search, which will lead to an error condition. Intentionally seeding errors into the action sequence enables the evaluation of error conditions that users are likely to encounter.

The last task-related preparation requirement is to ***identify one or more domain experts who will be available during the evaluation phase***. CC inspections are performed by a team of usability professionals, and it is unlikely that they will have expertise with the complex system being evaluated. Studies have shown that pairing domain experts with the team during the evaluation results in more effective results from usability testing of complex systems [48, 5]. A domain expert has a nuanced understanding of the intricacies of the domain and can provide insight into what the user is trying to accomplish and what the user would expect to experience within the task context. Such knowledge takes years to develop and cannot be sufficiently conveyed through short-term training.

### 4.1.3 Collaborative Critique Method-Related

An ***electronic spreadsheet***, in which evaluators will record answers to each of the questions in the CC inspection for each action/subaction in the task, is required. This spreadsheet, henceforth referred to as the **CC template**, must include the sequence of correct actions and subactions by subtask along one axis and the CC questions along the other. Since questions five through seven will only be answered when an error condition occurs, as noted in Section 3.1, values of "NA" (Not Applicable) can be predefined in the corresponding cells in the template, requiring evaluators to replace them with appropriate entries as needed.

## 4.2 Evaluation

The team of usability professionals, augmented by domain experts, perform the six-step evaluation procedure shown in Figure 4. Throughout the evaluation, they must keep in mind the knowledge,

1. Attempt to discern the next action to be performed based on the information in the task description and exploration of the interface.
2. Check the next (sub)action listed in the sequence of actions to ascertain that the walkthrough is proceeding correctly.
3. Perform that (sub)action.
4. Record in the template answers to the CC questions, which will be "Yes," "No," or "NA" (Not Applicable) in most cases; the exceptions are questions 2a and 2b, which require a numeric entry based on a predefined scale.
   - If the answer to a question is "No," then an explanation must be provided. Explanations may also be provided with other answers as deemed appropriate by the team but are not required.
5. If an error is reported by the system, answers to questions 5 through 7 must be recorded in the template. In all other cases, the answers to these questions will be "NA."
6. Repeat from step 1 until all of the actions specified in the sequence of correct actions have been performed.

Figure 4: Collaborative critique procedure.

experience, and skills a user can be expected to bring to the task. While the procedure is very similar to that of other walkthrough methods, one notable difference is that evaluators are encouraged to explore the interface when answering the CC questions. The reason for this is to enable them to judge the cognitive and physical efforts that would be expended by the user.

## 5 CC Walkthrough Example

To further illustrate the CC method, we present a sample evaluation consisting of a small portion of a Purchase Order (PO) task with an ERP system. While it is based on interactions with the actual system, we have simplified the action sequence to make it easier to follow. The background information included below has been greatly reduced from what would be provided to an actual evaluation team performing the entire PO task. The subtask should be understandable without training while still providing sufficient detail for demonstrating the scope of the CC questions.

Following the background information is a step-by-step CC evaluation of three actions in the subtask, with the evaluator's responses to the CC questions in italics. The questions are shown in abbreviated form (see Section 3.1 for the full versions). Only those questions to which the answers are either numeric or require explanatory text, as opposed to a simple affirmative or a Not Applicable response, are shown for each action. They are followed by comments that briefly analyze the evaluator's responses in the context of system-user collaboration.

---

**Background Information**

**Task:** Purchase Order (PO)

**Description:** A PO is defined as a request from the buyer (i.e., the purchasing organization) to a vendor for the purchase of materials and/or services. A purchase transaction is generated by creating and submitting a purchase order.

**Subtask:** Start a new PO and provide information to some of its header fields.

**Context description:** Generating a PO is part of the purchasing process, which includes procurement of materials and services from vendors and monitoring of deliveries and payments.

**User description:** The user is part of the purchasing organization and is responsible for the creation and submission of purchase orders. The user has been recently assigned to the purchasing team and has received basic training. She does know how to navigate the system's main menu but has no practical experience with the purchasing task and must be provided with any codes required for completing that task. A system manual is not available to the user.

---

**Collaborative Critique Evaluation**

**Action 1: Select the purchase order creation task.** In the system menu, select: Logistics - Materials Management - Purchasing - Purchase Order - Create - Vendor Known

**Q1. User finding options for what he wants to do in current screen:** *No - difficult to navigate due to so many levels in the menu structure and unhelpful labels.*

**Q2a. Exploration:** *4*

**Q2b. Confusion:** *4*

**Q3a. System guiding user based on knowledge:** *No - there doesn't appear to be a list of favorite or frequently performed tasks.*

**Q3b. System reducing user effort based on knowledge:** *No - there is no search by code or description, no obvious way to add frequently performed tasks to favorites, and way too many options in the menu. Should be filtered to show only those options that are relevant to the user.*

**Comments**: The evaluator's responses point to a lack of collaborative support on the part of the system due to its failure to effectively communicate the relevant options to the user and/or provide obvious means for the user to communicate her needs and discover the correct option. Providing search options and easily configurable favorite lists, as suggested by the evaluator, would reduce the effort required of the user.

The system must also use its knowledge of the user, including any available information on the user's role and past interactions with the system, to provide appropriate guidance. If such information is not available or helpful, then the system must provide a way for the user to be able to communicate her needs so that the system can be responsive.

The system must also pay attention to the actions of the user and respond accordingly. It may have been possible in this particular case for the system to detect the user's confusion based on, for example, a greater than expected number of menu interactions. The system should then offer help, such as directing the user to an appropriate search interface.

**Action 2: Specify the vendor for the purchase order.** Enter 1711 into the Vendor field.

**Q1. User finding options for what he wants to do in current screen:** *No - thought the correct field was the one immediately to the left of the Vendor label, not the one to the far right. Eventually figured this out after getting an error message.*

**Q2a. Exploration:** *3*

**Q2b. Confusion:** *4*

**Q3a. System guiding user based on knowledge:** *No - this is not the first input field and it is not clearly labeled.*

**Q4b. User understanding what remains to be done to complete overall task after performing action:** *No - after finally specifying the vendor correctly, there was a beeping noise and a warning message popped up saying to "Enter Purch. Org." That must be the next step, but there is no way to tell at this point what other information will be required. Rather than putting up annoying messages, it would be more helpful if all the required fields were marked as such and the cursor moved to the next required field upon the user pressing the tab key.*

**Q5. System displaying information clearly explaining error situation problem to user.** *No - entered 1711 into wrong field at first and got message stating "Document number 1711 not within defined interval." I did not understand what that meant, and I don't believe it would be at all clear to a user who is new to purchasing. It wasn't at all clear to me that I was just in the wrong field.*

**Q6. System presenting steps user can take for possible corrective actions:** *No. If the system had presented possible actions, then I would likely have realized sooner that I was not in the vendor field. If the user had actually needed to fill in this unlabeled field, there was no information on what is the "defined interval" that the input needs to be within.*

**Q7. System presenting easy way to take corrective actions:** *No. If the system had presented a list of appropriate values for the field, I would have known I was in the wrong* *field, and the user who meant to be here would have access to valid responses.*

**Comments**: The evaluator was confused by the placement of fields and labels in the PO interface, which has an unlabeled Document Number text field, a Vendor label adjacent to that field, a considerable amount of space after the label, and then a Vendor text field. Given this layout, it is not surprising that the evaluator entered the vendor identifier of 1711 into the Document Number field instead of the Vendor field, as she had intended. Rather than misleading the user, the system must make the informational data requirements for each field obvious. In this case, the first field should have been clearly labeled as Document Number and the Vendor label should have been adjacent to the Vendor field.

The requirement to share relevant information also means that required fields must be clearly indicated. To function as a helpful partner to its users, the system should limit its use of disruptions such as beeping messages to only those contexts where they are absolutely necessary.

Another requirement implied by the commitment to the success of the joint activity is that the system must match its vocabulary to that of the user. The error message was confusing because there was no explanation given for what the defined interval is for the document number or any suggestions for valid responses.

**Action 3: Specify the purchasing organization.** Select 3012 from the drop-down list of available purchasing organizations.

**Q1. User finding options for what he wants to do in current screen:** *Yes - the option for specifying the purchasing organization is obvious. I ran into difficulties after this, however, and did not find options for taking corrective action.*

**Q2a. Exploration:** *1*

**Q2b. Confusion:** *1 - in terms of being able to select 3012. Then after I got an error message, I became very confused.*

**Q3a. System guiding user based on knowledge:** *No - I found 3012 easily enough, but once I selected it I got an error message stating that "Vendor 1711 had not been created for purchasing organization 3012." The purchasing org shouldn't have been included in the list if it wasn't a valid selection given my other input.*

**Q4a. User understanding progress so far toward completing overall task after performing action:** *No - the error message makes me unsure of what progress was made.*

**Q4b. User understanding what remains to be done to complete overall task after performing action:** *No - it is not clear what I will have to do to get back on track.*

**Q5. System displaying information clearly explaining error situation problem to user:** *No - the vendor is in the system, so is the problem really that it hasn't been created yet or is it just that it hasn't been assigned to the purchasing organization?*

**Q6. System presenting steps user can take for possible corrective actions:** *No - the system hasn't presented me with any options, and there are far to many tabs and choices for someone new to the purchasing process to be able to figure out on her own what to do next.*

**Q7. System presenting easy way to take corrective actions:** *No - there is no indication of where to go to create the vendor for this organization.*

**Comments**: In this ERP system, the purchasing organization can only buy from a vendor that has been added to the list of vendors associated with that organization. The evaluator ran into a problem because the vendor she had specified in Action 2 (i.e., 1711) had not yet been added to the purchasing organization selected in this step (i.e., 3012). As the evaluator pointed out, the system should present valid options to the user rather than ones that will result in errors if chosen. The selection of the purchasing organization in the above case was not necessarily invalid, assuming additional actions can be taken to add vendor 1711 to that organization. The system has the knowledge of whether or not such an action is supported by its functionality, is in accordance with the implemented business rules, and can be performed by the user based on the authorization level requirements.

Instead of displaying an error message, the system should either have not included organization 3012 as a possible option if it is truly an invalid choice, or it should have communicated what the problem is and presented the possible action(s) that the user can pursue to achieve the current goal. For example, it could provide a means for adding vendor 1711 to purchasing organization 3012 if that action is allowed for the user. If it is not allowed, it could present a list of vendors that have been added to the purchasing organization so that a different vendor could be selected.

# 6 Evaluation of the CC Method

In this section, we highlight selected findings from an initial evaluation of the CC method and consider them in light of other reported results regarding CC and other inspection methods.

In our study ([11]), four two-person teams of usability professionals evaluated a leading ERP system in a laboratory setting using CC. Each of the evaluators received in-depth training in the task that was used in their evaluation. In addition, at least one of the evaluators per team had sufficient experience to qualify him or her as having domain expertise. The teams' responses to the CC questions were compared to the usability problems identified separately via user testing with 10 subjects. Two of the teams evaluated a typical administrative task interface; the other two evaluated a transactional task.

Problems identified in [11] via user testing were referred to as *usability problem instances (P Instances)*. Problem instances were also classified by *problem type (P Type)*, with repeated instances of the same problem considered to be the same type. Table 1 presents a sampling of the types of problems that were predicted within evaluator responses to particular CC questions. Notably, the vast majority of such predictions were made in response to more than one question corresponding to a step in the walkthrough, like in the example evaluation we presented in Section 5. Most of the predicted problems were ranked with a number greater than one in the answers to question 2 on user confusion and exploration. Evaluators often expanded on the reasons for the anticipated user confusion and exploration in their answers to questions 1, 3, and 4, and sometimes suggested potential solutions to an issue. There was no clear correlation between the type of usability problem and a specific question. For example, the *Main menu* problem, describing the difficulties users experience in locating a desired transaction in the main system menu, is referred to in answers to questions 1 and 2 in one case, and question 3 in another.

One of the ways of assessing a usability inspection's effectiveness involves computing the method's *thoroughness* and *validity*, as defined by Sears [35]. Thoroughness refers to the proportion of real problems found with the method to real problems that exist; validity is defined as the proportion of the number of real problems identified by the method to the number of all problems, including "false positives," predicted by the method. As we noted in Section 2, a direct comparison of these effectiveness measures demonstrated in various studies is acknowledged to be problematic ([49, 24]). However, in Table 2 we present a set of results reported in several studies to illustrate the range of values.

Table 2 entries include results of individuals studies with the exception of the last row, which presents a summary of seven different studies. Entries in the table containing a range of thoroughness and validity values correspond to the results from multiple evaluator teams. The effectiveness metrics of CC reported in our initial evaluation (row 1) are in line with another independent assessment by Zaman [50]. We have not found any studies that report on the thoroughness or validity of MOT, but a very comprehensive experimental evaluation of MOT is presented in [37].

Another important metric for usability evaluation methods is the *severity ratings* of the usability problems identified by the method, since problems with high severity levels are considered to be more important to find and fix than those with lower levels [49]. In our evaluation of the CC method, each of the problem instances was rated on a severity scale ranging from 1 for least severe to 3 for most severe. These ratings were derived from the results of user testing and were based on user-reported confusion and frustration levels and the parameters observed via an inspection of the screen capture of the user performing the task. The observed parameters included exploration time, success in completing the step, and the effect of the problem on the task outcome. The precise factors determining problem instance severity are presented in Table 3. To summarize, problem instances were categorized as the least severe if they occurred in a step that the user was able to complete with either no or little exploration and no or mild confusion, or with moderate exploration that might have left the user mildly confused and frustrated. We assigned the highest severity rating to a problem instance if it had a negative effect on the overall task outcome, or if it occurred in a step that either the user was unable to complete successfully after spending more than two minutes exploring the interface or resulted in a combination of high confusion, high frustration and long exploration time. All other problem instances were assigned the moderate severity rating.

| Question numbers | Problem type |
|---|---|
| 1 | Inconsistency in interface (e.g. Checkmark vs Save icon; what is clickable and what isn't) |
| | Layout of label and field (misleading layout causing user error ) |
| 1;2 | Main menu (difficulty locating a transaction in the main menu) |
| | Long exploration (User spent more than 25 seconds exploring the interface before proceeding to take a step) |
| | Layout of label and field (misleading layout causing user error ) |
| 1;2;3 | Unclear that Detail section has to be expanded |
| | Difficult to work with codes |
| | User selects wrong option |
| 1;2;3;4 | Confusion over Create PO icon |
| | Confusion about the next step |
| | Single Role button (user didn't know needed to press this button) |
| 2 | Difficulty finding an entry in the possible values |
| | Hard to find relevant column |
| 2-3 | Confusion over sequence of actions in authorization process |
| | Not clear where to enter amount of discount |
| | Unhelpful default placement of cursor |
| 3 | Main menu (difficulty locating a transaction in the main menu) |
| | Confusion about the next step |
| | Currency field not populated automatically |
| 4 | Unclear about the purpose and content of the current interface |
| | Confused about where you are in the process |
| | Skipped step |
| 5 | Difficult to work with codes (Referring to the codes that appear in the error message) |
| 5;6 | Incomprehensible error message |
| 5;6;7 | Confusion over warning message concerning record flagged for deletion |
| | Did not look up plant for material |
| 7 | Insufficient guidance with error message |

Table 1: A sampling of problem types predicted in the responses to CC questions.

The evaluation showed a relatively uniform distribution of CC predictions among the problems of varying severity: 52%, 60%, and 54%, of problems with severity 1 through 3, respectively, were predicted for one of the tasks, and 79%, 89%, and 78%, respectively, for the other task.The relative uniformity of prediction rates across all severity levels indicates that the CC method does not suffer from the shortcoming commonly reported with other inspection methods of detecting a larger portion of trivial problems. Regarding the differences in prediction rates between the two tasks, we believe that the higher numbers have more to do with the experience of the evaluators rather than the nature of the task. Similar differences between novice and expert evaluators' CC results are reported in [50].

| Study | Thoroughness | Validity | Domain/Notes |
|---|---|---|---|
| Babaian et al., 2012 [11] | CC: 39-67% P Instances<br>47-68% P Types | not assessed;<br>false positives in 5.9% of 118 steps of all walkthroughs | ERP interface tasks, expert evaluators |
| Zaman, 2012 [50] | CC: 47-71% (expert)<br>14-46% (novice) | not reported | Websites, expert and novice evaluators |
| Koutsabasis et al., 2007 [51] | HE: 20-24.3%<br>CW: 24.3-25.7% | HE: 60.7-100%<br>CW: 70.8-85.7% | Website, novice evaluators |
| Chattratichart & Brodie, 2004 [52] | HE: 65%<br>HE-Plus*: 76% | HE: 28%<br>HE-Plus: 39% | Website, novice evaluators (*w/problem profile) |
| Hvannberg et al., 2007 [53] | HE: 36% | HE: 40% | Web portal |
| Hartson et al., 2003 [49] | HE: 31-86%<br>CW: 15-17% | HE: 17-46%<br>CW: 5-73% | Summary of data from 7 studies comparing HE to other methods |

Table 2: Usability method studies reporting thoroughness and validity.

| | |
|---|---|
| Severity 1 (low) | Step completed AND<br>(confusion <=2 AND exploration <30 seconds OR confusion <=2 AND frustration <=2 and 30 seconds <exploration <60 seconds) |
| Severity 2 (moderate) | all others |
| Severity 3 (high) | Negative effect on task outcome OR<br>Non-completion of step AND exploration >120 seconds OR<br>(confusion >=3 OR frustration >= 3 ) AND exploration >120 seconds OR<br>confusion >=3 AND frustration >= 3 |

Table 3: Factors determining problem severity levels. *Confusion* and *Frustration* levels are self-reported by users and are as follows: 1 - none, 2 - somewhat, 3 - very, 4 - extremely.

For complex systems, the ability to *detect error conditions resulting in usability problems* is also an important metric. The results of the evaluation confirmed the strength of the CC method in predicting usability problems associated with error situations, as 96% of all such problem instances and 93% of problem types were identified by the walkthrough responses.

# 7 Discussion

The CC method is conceptually different from other inspection techniques in that it explicitly focuses on the helpfulness of the system in the process of working with the user on achieving a goal. The main emphasis of this method is on reducing the physical and mental efforts expended by the user.

To achieve this requires paying explicit attention to the task, the user, and the interaction contexts in designing system-user interaction and communication. We have chosen to ground the method in collaboration theory because it directly addresses the phenomena that distinguish merely working together from working on a common goal with a commitment to maximizing the chances of its successful achievement [14]. Other theories and computing methodologies, such as activity theory [54], distributed cognition [55], situation awareness [46, 56], and collaborative planning [15], can be used to further elucidate different components of interface design in order to best meet the requirements of collaboration. They can help answer questions about *how* to make the system behave more collaboratively, such as how to partition the task optimally, what contextual information to include in communications, and when to interrupt the user to ask a question or deliver important information. Collaboration theory, however, provides useful guidance in focusing design methods on *what* the overall interaction design should accomplish.

A distinguishing feature of the CC method is that the expert assessment must take into account not only the descriptions of the user and the user interface but also the knowledge, functional capabilities, and behavior of the system as related to a particular task. These unique aspects have important implications for both evaluating interactions and guiding redesign efforts. Frequently noted limitations of inspection methods ([12]) include (1) a tendency to focus on specific user difficulties, (2) no identification of missing functionality, and (3) discovery of a greater number of relatively insignificant problems while missing more severe ones due to insufficient consideration of task context. The conceptual design of the CC method intentionally does not restrict the problem space to any predefined set of usability issues in order to address the first two shortcomings; the true efficacy of the method with respect to these two issues, however, needs to be verified in practice. The evaluations presented in [11, 50] and summarized in Section 6 suggest that this method does not suffer from the third shortcoming, with relatively uniform prediction rates across all severity levels.

In comparing CC with CW, the cognitive walkthough method also relies on four questions (see Figure 5), but its focus is on ease of learning [27], while CC focuses on the helpfulness and usefulness of a system in reducing the user's physical efforts and cognitive load. There are some similarities in intent between some of the questions in each method:

- CC question 1, which explores the match between the user's mental model of what needs to be accomplished and its representation in the system, roughly corresponds to CW questions 1, 2, and 3.

- CC question 2 digs deeper than CW questions 2 and 3 by explicitly assessing the need for (a) exploration and (b) the amount of user confusion in finding the correct action to take.

- CC question 3 goes beyond CW question 3 by emphasizing the system's ability to share knowledge and lessen the load on the user.

| |
|---|
| 1. Will the user be trying to achieve the right effect? |
| 2. Will the user notice that the correct action is available? |
| 3. Will the user associate the correct action with the desired effect? |
| 4. If the correct action is performed, will the user see that progress is being made? |

Figure 5: Four cognitive walkthrough questions.

- CC question 4a corresponds to CW Question 4

There are no counterparts to CC questions 4b, 5, 6, or 7 in the CW questions.

Another key difference between the two methods is in their implementation. CC evaluators are directed to explore the interface before proceeding with the specified correct action in questions 2a and 2b. This opens up a CC evaluation to potentially discovering a wider range of issues than would be uncovered with strict adherence to the action sequence, as imposed by CW, and is particularly important for evaluating complex systems requiring the integrated use of multiple system features. While the heuristic walkthrough method supports exploration and is task-focused (from its heuristic evaluation and CW forebears, respectively), the CC questions overall are more firmly rooted in complex task performance than either the CW questions or Nielsen's heuristics.

A study by Zhang et al. [57] indicated that the predictive power of a usability inspection is increased by focusing an evaluation on three usability perspectives: novice use, expert use, and error handling. Similarly, the CC method includes implicit differentiation of user categories in Question 2 and also places particular emphasis on evaluating system performance in error situations (Questions 5 – 7). In an HW, error situations are evaluated in light of Nielsen's heuristics on error prevention and helping users recognize, diagnose, and recover from errors (see Figure 1 and [33]). The three CC questions focused on error conditions include a stronger emphasis on the system's role in developing and executing an error-handling plan. They extend the requirement on the active usage of task and interaction context to reducing the user's workload in error situations, which are known to cause some of the greatest obstacles to productivity.

Like the metaphors of thinking (MOT) method, CC evaluations do consider the user's thinking by assessing confusion (Question 2b) and understanding (Questions 4a and 4b) within the specific context of use, evaluating if the user will find the correct option (Question 1), and estimating the amount of exploration required by users with a range of experiences (Question 2a). These questions involve less of a psychological basis than MOT but are more clearly centered on usability. Like MOT and CW, CC is also not limited to particular devices/interactions due to its emphasis on the collaborative behavior of the system. Given the similarity of MOT to HE, we conjecture that collaborative critique, with its underpinnings in collaboration theory, explicit focus on error situations, and consideration of task, user and interaction contexts, is better suited than other methods to evaluation of complex systems. This supposition must, of course, be evaluated in practice.

## 7.1 Limitations

Although its name may suggest otherwise, the CC method in its current form is most suited to evaluating interactions between a system and a user, as opposed to evaluating collaborations between multiple users of a system. While inspection methods can be applied early in the development process [30], some of the features that serve to differentiate the CC method also make it more suitable for evaluating an implemented prototype. For example, explorations of the interface and evaluations of the system's responsiveness in error situations are most fully realized when the evaluator can actively interact with a prototype, encounter errors while doing so, and experience the system's responses. At the same time, CC can serve as a useful guide in directing early design efforts by emphasizing the need for maintaining the context of the interaction while working with the user.

Inspection methods are also noted for being less expensive due to their reliance on experts rather than users, but they are inherently expensive in the amount of time it takes for an evaluation to be performed. The CC method is no exception, with the teams in [11] allotted three hours for the inspection and one of the teams requiring additional time. Having one evaluator act as the navigator while the other enters responses to the CC questions helps to improve the efficiency of the process and lessen the burden on each evaluator, but it may also lengthen the evaluation time, as it encourages discussion between the team members. The fruitfulness of such discussions is deemed to be more important to the effectiveness of an evaluation than the tradeoff with time.

Most of the CC questions require that the evaluators consider themselves in the role of a particular type of user. This is particularly necessary in evaluating complex systems, in which a user's role and experience define the scope of her knowledge of task parameters and may dictate restrictions on access. Questions 2a and 2b, however, ask the evaluators to consider users with a range of experience levels in estimating the amount of exploration and confusion that is likely to be involved. Since no description of what that range may look like is provided, these questions are more difficult for evaluators to answer. Minor adjustments to these questions and the background information provided to the evaluators are needed to address this issue. Larger changes would be required for modifying the CC questions to accommodate the evaluation of collaborative tools by considering the perspectives of a group of users working together to accomplish a goal.

The quality of the results achieved with a walkthrough evaluation is affected by a number of factors related to the application of the method. The parameters of the task being evaluated are likely to have some influence on the outcome. For example, if a task is more engaging to the evaluators, they may be more likely to spend time exploring the interface. The experience of the evaluators also affects how well the method performs, as confirmed in [11, 50].

Another issue related to the performance of the CC method was that the teams in the study were not asked to derive lists of usability issues, which limited the way in which false positives could be evaluated and likely led to underestimations of those values. In addition, as noted by [58], usability evaluation methods should lead to design changes that correct the identified problems. Significant efforts must be devoted to working closely with system designers in order to derive actionable design solutions based on the results of an evaluation.

# 8 Conclusions

We have presented the collaborative critique usability inspection method, which is differentiated by its focus on assessing the collaborative strength of a system. The CC questions direct attention to the usefulness of the system in light of user, task, and interaction contexts. The method is further distinguished by the inclusion of questions specific to error situations, which often have a critical impact on system usability, and the encouragement of interface exploration throughout the evaluation process for promoting the user's perspective.

We have described the CC walkthrough procedure in detail, including task selection, material preparation, and the evaluation process, the latter of which was illustrated with an example of a walkthrough inspection. That example demonstrates how evaluating system-user interactions from a collaborative perspective highlights the ability of the system to act as a responsive partner that is committed to helping its users achieve their goals. This ability becomes increasingly important with increasing system size and complexity, as the user can only be expected to be familiar with a subset of the available functionality and should be able to rely on the system for guidance and support. The explicit focus on error situations in the CC questions underscores the importance of evaluating how the system behaves and responds when users are most likely to be in need of assistance.

In the future, we will be evaluating the CC inspection process itself, with the goal of optimizing the performance evaluators can achieve using this method. Such factors as the number of evaluators that should ideally participate in a walkthrough and the amount of training they should receive in collaboration theory and the CC method will be under investigation. We will also be developing additional task scenarios for use in studies in both the laboratory and the field aimed at further assessing the strengths and weaknesses of this method.

The ultimate test of the CC method will be its effectiveness in guiding developers toward more collaborative design solutions that improve the performance and usability of complex systems. Field evaluations will allow us to investigate issues involved in applying the outcomes of CC inspections. Of primary concern will be in what form findings should be conveyed for encouraging dialogs that focus on system-user collaboration as a critical component in each stage of the design process.